\documentclass[twocolumn,showpacs,preprintnumbers,amsmath,amssymb]{revtex4}

\usepackage{graphicx}
\usepackage{dcolumn}
\usepackage{bm}

\begin{document}

\preprint{APS/123-QED}

\title{LiBC - A prevented superconductor}
\author{B. Renker,$^{1}$ H. Schober,$^{2}$ P. Adelmann,$^{1}$ P. Schweiss,$^{1}$
       K.-P. Bohnen,$^{1}$ and R. Heid$^{1}$}
\affiliation{$^{1}$Forschungszentrum Karlsruhe, IFP, P.O.B. 3640,
       D-76021 Karlsruhe, Germany}
\affiliation{$^{2}$Institut Laue-Langevin, BP 156 X, F-38042
       Grenoble Cedex, France }

\date{\today}

\begin{abstract}
We have investigated the lattice dynamics of LiBC. Experimental
Raman and inelastic neutron scattering results are confronted
with density functional calculations. The excellent agreement
between experiment and theory indicates that we have attained a
high level of understanding for LiBC, which by its structure is
closely related to superconducting MgB$_2$. We will show that
annealing of the sample is insufficient to cause Li deficiency
and thus produce the expected superconductivity.

\end{abstract}

\maketitle

The simple binary compound MgB$_2$ has attracted considerable
attention because of its exceptional superconducting properties.
A strong coupling of the electronic system to high-frequency
boron in-plane phonon modes meets the BCS requirements for a
transition temperature as high as 40\,K \cite{Kortus,Bohnen}. A
most prominent experimental manifestation of this coupling is the
extreme broadening of the E$_{2g}$ mode in the Raman spectrum
\cite{Bohnen,Renker}. Very recently theoretical speculations of
transition temperatures as high as 120\,K i.e. largely exceeding
those of MgB$_{2}$, have been published for LiBC compounds with
Li deficiencies in the range of 25\,\% \cite{Rosner}. With respect
to structural peculiarities LiBC provides a comparable system to
MgB$_{2}$. B and C atoms alternately occupy the sites within the
hexagonal sheets. Charge neutrality is restored by replacing
Mg$^{+2}$ by Li$^{+}$. A doubling of the unit cell along the
hexagonal axis occurs due to an interchange of B and C atom
positions in neighboring planes\cite{Wörle}. So far no convincing
experimental evidence for superconductivity could be established
\cite{Bharathi,Souptel}. In this context, it is most interesting
to investigate the lattice dynamics of this particular compound
and to compare it with our previous results for MgB$_{2}$ and
AlB$_2$.

LiBC has been synthesized following a procedure which is
described in detail in Ref. \cite{Wörle}. In view of our neutron
scattering experiments we have used isotope pure $^{7}$Li metal
together with amorphous $^{11}$B and graphite powders. The mixed
components were pressed into pellets and held in a closed Nb tube
at 1500$^{\circ}$C for 3\,h. The final product was homogeneous
and consisted of yellow shining crystallites with clean faces and
metallic brightness. Stoichiometric LiBC has been shown to be
semiconducting and to belong to the space group P6$_3$/mmc which
was also confirmed by our own $x$-ray refinements. Hexagonal
shaped platelets with diameters of up to 200\,$\mu$m and
$20\,\mu$m in thickness could be found and were chosen for
optical measurements. All of our samples - the as prepared, the
annealed and the Na doped ones - proved to be non-superconducting
above 4\,K. For inelastic neutron measurements the
microcrystalline powders were filled into flat pockets of
$2\times\,3\times\,0.3$\,cm$^{3}$ made from Al-foil. Measurements
were performed at 300\,K on the time-of-flight spectrometers IN6
(incoming energy 4.7\,meV) and IN4 (incoming energy 39.4\,meV) at
the HFR in Grenoble (France) in the upscattering mode. A
generalized phonon density of states (GDOS) was obtained from the
sum of scattering spectra recorded over a large angular range from
10$^{\circ}$ to about 120$^{\circ}$. The IN6 and IN4 GDOS
conincide well above 10\,meV indicating that the incoherent
approximation works very well. Only small coherence effects are
observed in the IN6 data below 10\,meV due to the fact that the
$Q$-space is less extensively sampled with the longer wavelength
neutrons at these small energy transfers. Multi-phonon
contributions were removed via a self-consistant calculation
procedure.

The factor group analysis yields 2A$_{2u}$ +2B$_{1g}$ + B$_{2u}$
+ 2E$_{1u}$ + 2E$_{2g}$ + E$_u$ zone center optic modes where
only the 2E$_{2g}$ modes are Raman (R) active. The latter modes
should be observable for all configurations with polarization
vectors of the incoming and the reflected light within the
hexagonal plane. Zero intensity is expected if one of these
vectors is parallel to the crystallographic $c$ axis. The spectra
shown in Fig.\,1 prove that this is essentially the case. In
particular we do not observe additional strong and sharp lines at
546\,cm$^{-1}$ and 830 cm$^{-1}$ which have been reported by
other authors \cite{Hlinka} and which have been taken as a reason
to propose the lower symmetric space group $P\overline{3}m1$ for
LiBC.
\begin{figure}
\includegraphics[width=0.9\linewidth]{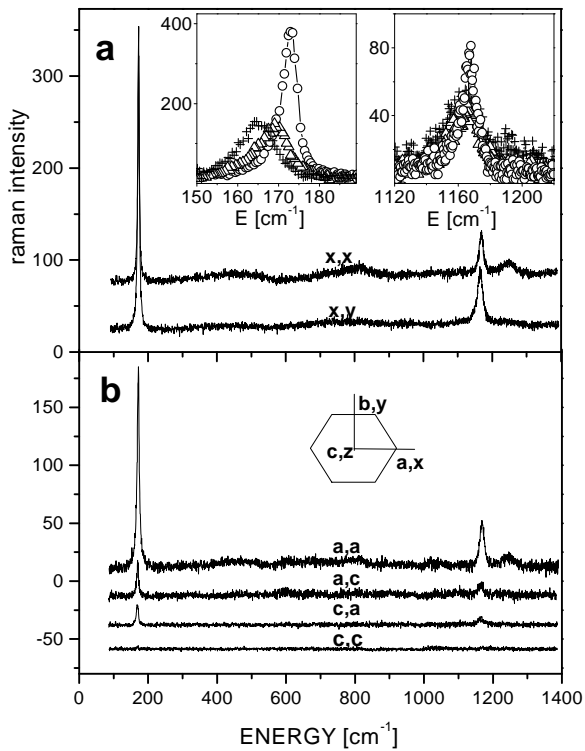}
\caption{Observed Raman signals from a LiBC single crystal in
        backscattering configuration. a: scattering from the hexagonal
        plane, the small contribution near 1240\,cm$^{-1}$ in the
        polarized spectrum is not allowed for P6$_3$/mmc symmetry.
        Results for the treated materials are shown by inserts: open
        circles - as prepared, triangles - annealed at 1000$^{\circ}$C,
        crosses - Li$_{0.95}$Na$_{0.05}$BC, b: scattering from an
        ac-plane: the lettering above the spectra marks the
        polarization vectors of the incoming and outgoing light. In
        agreement to the selection rules strong Raman signals are only
        observed if both polarization vectors are contained within the
        hexagonal plane.}
        \label{fig1.eps}
\end{figure}
These authors conclude on a particular ``puckering" of hexagonal
B-C planes. We have never observed these lines although we have
noticed that the background spectrum may depend on details of the
sample preparation. A high Li vapor pressure at the synthesis
temperature around 1500$^{\circ}$C and possible small leaks in
the container may cause problems in sample preparation. However, a
broader contribution around 1270\,cm$^{-1}$  is also observed in
our polarized spectra. It is not allowed as a one phonon
contribution. Its intensity when compared to the allowed line at
1168\,cm$^{-1}$ is sample dependent and gets weaker on cooling.
It is attributed to second order scattering or to structural
defects although its origin is not fully understood.

The important hole doping of B $2p-\sigma$ bands in MgB$_2$ is
caused by the relative shift of $\pi$ and $\sigma$ bands in this
particular compound. This does not occur for LiBC. However, there
are speculations that a doping could be achieved by Li deficiency.
We have annealed samples at 800$^{\circ}$C, 1000$^{\circ}$C and
1200$^{\circ}$C for $\geq 3$\,h in dynamic vacuum in order to
reduce the Li content. Structural changes within the Li sheets
which provide a coupling between neighboring B-C planes are
expected to show up in the sliding frequency at 171\,cm$^{-1}$
(see Tab.\,1). However, little changes are observed in the
R-spectra up to annealing temperatures of about 800$^{\circ}$C
when Li starts to evaporate. For the sample which was annealed at
1000\,K we observe a small but significant downshift of the
sliding frequency indicating a possible decrease in the Li
concentration. A somewhat larger shift was observed for a sample
of nominal Li$_{0.95}$Na$_{0.05}$BC. For the B-C stretching mode
at 1167\,cm$^{-1}$ we register a still smaller relative shift
which let us conclude that the obtained changes in the electronic
system are negligible. These results were reproducible on
different crystallites and are shown in detail by the inserts in
Fig.\,1. Thus neither the proposed dramatic softening nor any
significant line broadening as observed for MgB$_2$ occur. In
view of the high frequency of the E$_{2g}$ in-plane mode LiBC
resembles much more the compound AlB$_2$. Indications of sample
disintegration, i.e. a decrease in intensity of the two E$_{2g}$
lines and broader contributions at energies $>$\,1200\,cm$^{-1}$,
are observed for samples with annealing temperatures
$>$\,1370\,K. Since samples which already start to show
signatures of disintegration in the R-spectrum still exhibit the
correct $x$-ray structure it is concluded that disintegration
starts on the surface of the crystallites.

\begin{figure}
\includegraphics[width=0.8\linewidth]{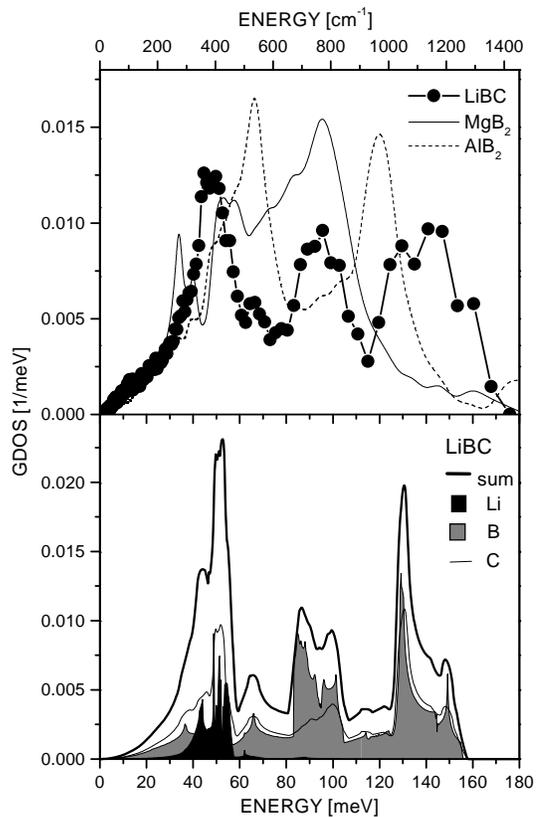}
\caption{Experimental generalized phonon density of states for as
prepared LiBC, superconducting MgB$_2$ and non-superconducting
AlB$_2$ (the latter two spectra are results from preceding
measurements \cite{Renker} under the same conditions). A
pronounced down-shift and broadening of the high energy peak
visible for MgB$_2$ is absent for LiBC. LiBC resembles much more
AlB$_2$. Lower frame: Theoretical results for stoichiometric LiBC
proving the high quality of calculations based on density
functional theory. Results for the partial DOS are helpful for a
more detailed interpretation of the experimental spectrum.}
\label{fig2.eps}
\end{figure}

Augmenting results are obtained by inelastic neutron scattering
which can access all excitations in the entire zone and which
investigates the whole volume of the crystallites. In Fig.\,2 we
show the generalized phonon density-of-states measured at 300\,K.
The spectrum consists of three main peaks. Hard B and C in-plane
and out-of-plane modes are grouped around 150\,meV and 100\,meV
respectively. Modes where neighboring B-C sheets slide against
each other as well as vibrations of the weaker bond Li-ions
contribute to the lowest peak around 50\,meV. Fig.\,1 shows that
the upper two peaks in LiBC group around the corresponding line
of AlB$_2$ which reflects the doubling of the unit cell in $c$
direction . A shoulder around 160\,meV agrees well to the
``forbidden" line at 1270\,cm$^{-1}$ in the R-spectrum. A clear
result is that the significant down-shift and broadening of the
high energy peak of MgB$_2$ does not occur for our sample of as
prepared LiBC. Therefore, the spectrum of undoped LiBC resembles
much more the GDOS of AlB$_2$. The exceptional behaviour of
MgB$_2$ is besides structural differences the result of strong
electron-phonon coupling for the E$_{2g}$ mode (1167\,cm$^{-1}$
for LiBC). Here, both experiments agree that nothing comparable
exists for the as prepared sample of LiBC. A hybridization of Li
modes with the underlying B(C) partial DOS is however also
registered in the GDOS spectra of LiBC.

A crucial question is, whether it is possible to obtain a
sufficiently strong doping of LiBC by Li evaporation. In
Ref.\cite{Wörle} the authors were confident that this could be
achieved. For superconductivity it would be important to obtain a
homogeneous doping of the entire material without damage of the
main structure. This however could not be observed in our
R-spectra which investigate the surface of the crystallites. With
INS we can see the whole volume. In Fig.\,3 we show GDOS spectra
which have been obtained for samples with different annealing
temperatures. It can be seen clearly that no shift in the peak
positions occurs. Up to $\approx$\,1120\,K the sample seems
stable. Annealing at still higher temperatures damages the whole
sample. The initially sharp peaks start to disappear remarkably
without any significant shifts in their position. A humb around
20\,meV rises up and signalizes the formation of amorphous
material. This spectral change is still more pronounced in the
original time-of-flight spectra. It is a clear result of our
measurements that not enough of the Li ions sandwiched between
the hexagonal B,C-planes can be removed in order to cause the
desired changes in the electronic system. Dramatic changes due to
an unusually strong electron-phonon coupling even more pronounced
than those found for MgB$_2$ otherwise should have been observed.

\begin{figure} [t!]
\includegraphics[width=0.90\linewidth]{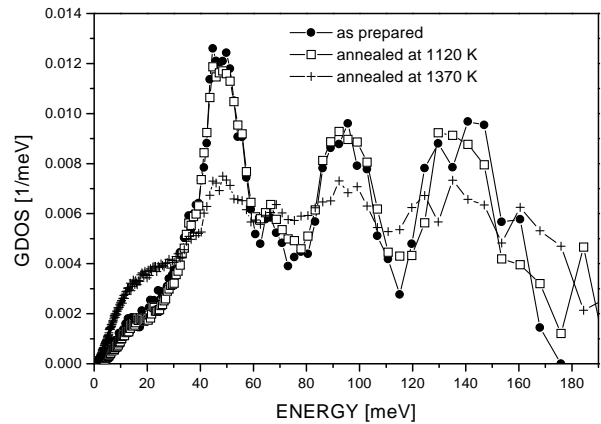}
\caption{Measured GDOS spectra for LiBC samples which were
annealed for 3\,h in dynamic vacuum at the indicated temperatures
in order to obtain a hole doping of B-C layers due to Li
deficiency. Changes in the electronic system are believed to
manifest in distinct peak shifts which are not observed even if
the sample starts to disintegrate.}
        \label{fig3.eps}
\end{figure}
\begin{table}
   \caption{Calculated $\Gamma$ point frequencies in meV (cm$^{-1}$)}
  \begin{ruledtabular}
   \begin{tabular}{c|c|c}
sym  & $\omega$ & elongations  \\
\hline
     E$_{2g}$ & 21.8    & E$_{2g}$, B-C planes slide against each other \\
     (R)      & (176)   & Li at rest \\
     E$_{2u}$ & 37.3    & Li layers slide against each other  \\
              & (301)   &  \\
     B$_{1g}$ & 39.5    & symmetric vibration of B-C layers \\
              & (319)   & along c \\
     E$_{1u}$ & 43.9 (T)& Li layers slide against B-C layers\\
     (IR)     & (354)   &  \\
              & 47.4 (L)&  \\
              & (382)   &  \\
     A$_{2u}$ & 56.7 (T)& Li layers vibrate against B-C layers  \\
     (IR)     & (457)   & along c \\
              & 69.9 (L)&  \\
              & (563)   &  \\
     B$_{2u}$ & 67.9    & Li layers vibrate against each other \\
              & (548)   & along c \\
     A$_{2u}$ &101.5 (T)& B and C layers move against each other \\
     (IR)     & (819)   & along c, Li not at rest  \\
              &104.2 (L)&  \\
              &(840)    &  \\
     B$_{1g}$ &104.5    & symmetric vibration of B-C layers \\
              & (843)   & along c, Li at rest  \\
     E$_{1u}$ &140.9 (T)& B-C bond stretching mode odd \\
     (IR)     &(1136)   & displacement of B-C layers  \\
              &152.7 (L)&  \\
              &(1231)   &  \\
     E$_{2g}$ &142.0    & B-C bond stretching mode even \\
     (R)      &(1145)   & displacement of B-C layers in phase  \\
       \end{tabular}
   \end{ruledtabular}
\end{table}
A detailed analysis of our experimental data has been performed
by calculations based on density functional theory. This
technique provides an accurate and parameter free analysis of the
electronic structure and the bonding properties. Our calculations
which correspond to those performed for other diborides
\cite{Bohnen,Heid} are based on a perturbational approach using
the mixed basis pseudopotential method and provide information on
the phonon frequencies and electron-phonon coupling within the
whole BZ. Pseudopotentials for Li and C \cite{Bohnen-Heid} were
constructed according to Hamann-Schl\"{u}ter-Chiang whereas for B
a Vanderbilt-type potential was chosen. Our calculations apply to
stoichiometric LiBC and the structure P6$_3$/mmc which assumes a
regular stacking of hexagonal B-C planes and sheets of Li ions
sitting on the corner of trigonal prisms. It can be seen from
Fig.\,2 that a very high level of correspondence to the
experimental GDOS is reached. All distinct maxima correspond to
similar features in the experimental spectrum. The very high
frequency shoulder at 160\,meV is not part of the calculated
spectrum which suggests an interpretation as a defect induced
contribution. Furthermore it can be seen from the partial DOS
that the Li modes overlap with B and C vibrations. Thus a
scenario similar to MgB$_2$ seems possible for samples with
sufficiently high doping. A significant feature of a very strong
electron-phonon coupling in MgB$_2$ was the drop of the E$_{2g}$
in-plane mode at $\Gamma$ well below the B$_{1g}$ out-of-plane
mode. In Tab.\,1 we have listed all calculated $\Gamma$-point
frequencies ordered according to increasing energies. For the
IR-active modes the LO\,(L) and TO\,(T) splitting has been
calculated. A good correspondence to the measured R-lines is
registered. Again with respect to the position of the E$_{2g}$
and B$_{1g}$ modes it can be seen that stoichiometric LiBC is far
away from the favourable situation found in MgB$_2$.

In conclusion, it is a success that larger single crystalline
specimens of LiBC with the shape of beautiful hexagonal prisms
can be prepared. We register that the R-selection rules are fully
obeyed for our samples. A small humb near 1270\,cm$^{-1}$ in the
(x,x)-spectrum probably originates from disorder. There is no
$x$-ray evidence of any second phase in the investigated single
crystals. Any significant doping by an evaporation of Li in
dynamic vacuum was not possible for the investigated single
crystals although very small frequency shifts occur for the
sliding mode at 171\,cm$^{-1}$. It is also a clear result that in
our preparation crystallites with a ``puckering" of the hexagonal
B-C planes were not obtained.

The generalized phonon density of states (GDOS) has been measured
by INS and proves the very high quality of first principals
calculations. The experimental spectrum of the undoped compound
resembles much more that of AlB$_2$ than MgB$_2$, i.e. signatures
of any stronger electron-phonon coupling are fully absent for
LiBC.

The preparation of doped LiBC crystals still remains a challenge
for sample preparation. Simple annealing in dynamic vacuum
however will cause a disintegration of LiBC crystals as
demonstrated in these investigations.

\end{document}